\documentstyle[12pt]{article}                                                                         
                                                                         
\newlength{\dinwidth}                                                                         
\newlength{\dinmargin}                                                                         
\def\lapproxeq{\lower .7ex\hbox{$\;\stackrel{\textstyle                                                                         
<}{\sim}\;$}}                                                                         
\def\gapproxeq{\lower .7ex\hbox{$\;\stackrel{\textstyle                                                                         
>}{\sim}\;$}}                                                                         
\def\be{\begin{equation}}                                                                         
\def\ee{\end{equation}}                                                                         
\def\bea{\begin{eqnarray}}                                                                         
\def\eea{\end{eqnarray}}

\begin{document}                                                                         
\titlepage                                                                         
\begin{flushright}                                                                         
DTP/98/98 \\                                                                         
December 1998 \\                                                                         
\end{flushright}                                                                         
                                                                         
\vspace*{2cm}                                                                         
                                                                         
\begin{center}                                                                         
{\Large \bf Penetration of the Earth by ultrahigh energy neutrinos} \\                                             
                                             
\vspace*{0.3cm}                                             
{\Large \bf predicted by low $x$ QCD}                                                                         
                                                                         
\vspace*{1cm}                                                                         
J.~Kwiecinski$^{a,b}$, A.D.~Martin$^b$ and A.M.~Stasto$^{a,b}$ \\                                                                         
                                                                        
\vspace*{0.5cm}                                                                        
$^a$ H.~Niewodniczanski Institute of Nuclear Physics, ul.~Radzikowskiego 152,                                                             
Krakow, Poland \\                                              
$^b$ Department of Physics, University of Durham, Durham, DH1 3LE \\                                                                        
\end{center}                                                                         
                                                                         
\vspace*{2cm}                                                                         
                                                                         
\begin{abstract}                                                                         
We calculate the cross sections for neutrino interactions with (isoscalar) nuclear                                              
targets in the energy domain all the way up to $10^{12}$~GeV.  Small $x$ QCD                                              
effects are included by using a unified BFKL/DGLAP formalism which embodies                                              
non-leading $\log 1/x$ contributions.  The few free parameters which specify the                                              
input parton distributions are determined by fitting to HERA deep inelastic data.  The                                              
attenuation of neutrinos transversing the Earth at different nadir angles is calculated                            
for a variety of energy spectra for neutrinos originating from different sources (from                            
Active Galactic Nuclei, Gamma ray bursts, top-down models), as well as for                            
atmospheric neutrinos.  For this purpose we solve the transport equation which                            
includes regeneration due to neutral current neutrino interactions, besides attenuation.                                              
\end{abstract}                                                                        
                                                                
\newpage                                                                        
\noindent {\large \bf 1.~~Introduction}                            
                           
The penetration of ultrahigh energy neutrinos through the Earth, with energies                            
$E$ greater than 10~TeV or so, can be strongly affected by neutrino                            
interactions with matter.  This is due to the increase of neutrino cross sections with                      
energy.  At these ultrahigh energies we have significant attenuation of the neutrino                      
fluxes on transversing through the Earth and, indeed, complete absorption for energies                      
above about $10^8$~GeV or so, depending on the nadir angle of the neutrino beam.                       
Realistic estimates of these effects are crucial for predicting the number of neutrinos                      
reaching the large km$^3$ scale detectors after penetration through the Earth.  Clearly                      
the magnitude of this ultrahigh energy neutrino flux has important implications for                      
neutrino astronomy.  For example, it is hoped that the neutrino flux coming through                      
the Earth will point back to its Active Galactic Nuclei (AGN) origin \cite{HALZEN}.                       
Clearly the magnitude of the signal, and the ability of the Earth to reduce the                      
background due to atmospheric neutrinos, are crucial in this exciting endeavour.                     
                           
The inelastic interaction of neutrinos with nucleons is traditionally described by the                            
QCD-improved parton model.   In ultrahigh energy neutrino interactions we are                      
probing a kinematical region which is not  accessible in current collider                            
experiments.  To be precise the most powerful electron-proton collider                            
HERA at the DESY laboratory in Hamburg typically probes the region                      
$x>10^{-4}$ for $Q^2 >10$~GeV$^2$, whereas the ultrahigh energy neutrino                      
interactions can become sensitive to the domain $x \sim 10^{-8}$ and $Q^2 \sim                      
M_{W}^2$.  Here, as usual, $x$  denotes the Bjorken scaling variable and $Q^2=-                     
q^2$,  where $q$ is the four momentum transfer between the leptons in the inelastic                      
scattering of a lepton on a nucleon.  The Bjorken variable $x$ is defined as                      
$x=Q^2/(2p.q)$ where $p$ denotes the four momentum of the nucleon.                            
                           
In perturbative QCD it is expected that the gluon and sea-quark distributions, and                      
hence also the deep inelastic scattering structure functions, should grow with the                      
decreasing values of the parameter $x$.  This theoretical expectation has been                 
beautifully confirmed by the structure function measurements at HERA \cite{DCSR}.                  
These measurements put important constraints on parton distributions in the small $x$                      
region probed at HERA.  In order to get predictions for the ultrahigh energy                 
neutrino-nucleon cross sections one has to construct a reliable extrapolation of the                 
structure functions to                      
the region of very small values of $x$ which is probed in these ultrahigh energy                      
interactions.  The existing extrapolations, which do also incorporate the constraints                      
from the HERA data, are based entirely on leading order (LO) or next-to-leading order                 
(NLO) DGLAP evolution                      
\cite{GQRS2,GQRS1,GKR}.  This approximation may, however, be incomplete at                      
low $x$ since it ignores the important resummation of $\ln(1/x)$ BFKL-type effects \cite{BFKL}.                  
In                      
this paper we wish to incorporate the QCD expectations at low $x$ which will take                      
these effects into account.  To be precise we shall base our calculation of the neutrino                      
cross sections on the unified BFKL/DGLAP formalism which incorporates both the                      
$\ln(1/x)$ resummation and the complete LO DGLAP evolution \cite{KMS}.  Due to               
the size of                      
the NLO $\ln (1/x)$ contributions \cite{NLO} one might question the reliability of                      
this procedure.  However our framework makes it possible to resum dominant                      
non-leading $\ln (1/x)$  contributions to all orders.  This has the important effect of                      
stabilizing the non-leading contribution and turns out to give a physically and                      
phenomenologically acceptable suppression of the LO BFKL behaviour.  In this way                      
we should achieve the most reliable dynamically-motivated extrapolation to very low                      
$x$ to date, which incorporates all the relevant QCD expectations.  The calculated                      
neutrino cross sections are then used as an input in  transport equation describing the                      
propagation of the neutrinos through the Earth \cite{NIK,NP}.   This equation will                 
contain                      
both the attenuation of the neutrino flux together with its regeneration through the                      
neutral current interaction. Similiar calculation which takes into account 
absorption and
gives prediction for the muon rates has been performed in \cite{HILL}.

The content of our paper is as follows.  In the next Section we discuss deep                            
inelastic lepton-nucleon scattering at low $x$ within the unified BFKL/DGLAP                            
framework.  In Sec. 3 we collect the relevant formulas needed for calculating the                            
neutrino cross sections and present our numerical results for these cross sections                             
calculated within the unified BFKL/DGLAP scheme. We also confront our                 
predictions                            
with results of calculations based on NLO DGLAP evolution.  Sec. 4 is devoted to                            
the discussion of the transport equation and its solution for the shadowing factor due                 
to passage of the neutrinos through the Earth.  In                            
Sec. 5 we present our results concerning the changes of the initial fluxes with nadir                 
angle.                             
We consider neutrino fluxes corresponding to Active Galactic Nuclei, Gamma Ray                            
Bursts and the Top-Down models, together with atmospheric neutrino background.                              
Finally in Sec. 6 we give our conclusions.   \\                                            
                                            
\noindent {\large \bf 2.~~Deep inelastic scattering at low $x$}                                             
                                            
Ultrahigh energy neutrino-nucleon interactions probe values of Bjorken $x$ which                                             
can be several orders of magnitude smaller than those which are accessible at the                                             
present deep inelastic $ep$ scattering experiments at HERA.  Here, as usual, $x =                                            
Q^2/2 p.q$ where $Q^2 \equiv -q^2$, with $p$ denoting the nucleon 4-momentum                                            
and $q$ being the 4-momentum transfer between the leptons in the deep-inelastic                                            
process $\ell N \rightarrow \ell^\prime X$.                                           
                                           
At low values of $x$ we must consider $\log (1/x)$ effects.  In ref.~\cite{KMS} a                                      
formalism is presented which permits an extrapolation of parton                                         
distributions to very small $x$.  Besides the leading order (LO) $\log (1/x)$                                         
resummation, the procedure incorporates a major part of an all-order resummation.                                          
We outline the method below.  We begin with the BFKL equation \cite{BFKL} for                                         
the unintegrated gluon distribution $f (x, k_T^2)$, which performs the LO $\log                                         
(1/x)$ resummation.  The equation has the form                                        
\be                                        
\label{eq:a1}                                              
f(x, k^2) \; = \; f^{(0)} (x, k^2) \: + \: \overline{\alpha}_S k^2 \int_x^1 \frac{dz}{z}                                         
\int  \frac{dk^{\prime 2}}{k^{\prime 2}} \left \{\frac{f(x/z, k^{\prime 2}) - f(x/z,                                         
k^2)}{| k^{\prime 2} - k^2 |} \: + \: \frac{f(x/z, k^2)}{[4k^{\prime 4} +                                         
k^4]^{\frac{1}{2}}} \right \}                                            
\ee                                        
where $\overline{\alpha}_S = N_c \alpha_S/\pi$ and $k = k_T, k^\prime =                                            
k_T^\prime$ denote the transverse momenta of the gluons, see Fig.~1.  The term in                                            
the integrand containing $f (x/z, k^{\prime 2})$ corresponds to real gluon emission,                                         
whereas the terms involving $f (x/z, k^2)$ represent the virtual contributions                                            
and lead to the Reggeization of the $t$-channel exchanged gluons.  The                                            
inhomogeneous driving term $f^{(0)}$ will be specified later.                                          
                                          
The observable nucleon structure functions $F_i$ are given in terms of the gluon                                           
distribution by the $k_T$ factorization formula \cite{KTFAC}                                           
\be                                          
\label{eq:a2}                                          
F_i (x, Q^2) \; = \; \int_x^1  \: \frac{dz}{z} \: \int \: \frac{dk^2}{k^2} \: F_i^{\rm                                           
box} (z, k^2, Q^2) \: f \left ( \frac{x}{z}, k^2 \right ),                                          
\ee                                          
where $F_i^{\rm box}$ describes the subprocess $Vg \rightarrow q\bar{q}$, see                                           
Fig.~1.  Here the virtual gauge boson $V$ may be either a $\gamma$ (describing an                                           
electromagnetic deep inelastic scattering) or a $W^\pm$ boson (describing a charged                                         
current weak interaction) or a $Z$ boson (describing a neutral current weak                                         
interaction).  The procedure (\ref{eq:a1}) and (\ref{eq:a2}) automatically                                           
resums all the leading $\log (1/x)$ contributions to the observable $F_i$.                                          
                                          
The solution of the LO BFKL equation for fixed $\alpha_S$ gives a QCD or hard                                         
pomeron with intercept $\alpha (0) = 1 + \lambda$ with $\lambda =                                         
\overline{\alpha}_S 4 \ln 2$.  The $\ln (1/x)$ resummation has recently been carried                                         
out \cite{NLO} at next-to-leading order (NLO).  It is found to give a very large $O                                         
(\alpha_S^2)$ correction to $\lambda$                                        
\be                                        
\label{eq:b2}                                        
\lambda \; \simeq \; \overline{\alpha}_S \: 4 \ln 2 (1 - 6 \overline{\alpha}_S),                                        
\ee                                        
which implies that the NLO approximation is unreliable for realistic values of                                         
$\alpha_S$. Rather we must use a formalism which contains an estimate of an                                         
all-order resummation.  Clearly it would be desirable to identify physical effects                                         
which could be resummed to all orders and which at the same time yield a NLO value                                         
of $\lambda$ that is comparable to (\ref{eq:b2}).  As it happens the imposition of the                                         
consistency constraint \cite{KMS2,AGS}                                          
\be                                          
\label{eq:a3}                                          
k^{\prime 2} \; < \; k^2/z                                          
\ee                                          
on the real gluon emission term gives just such an effect.  The variables are shown in                            
Fig.~1.  The origin of the constraint is the requirement that the virtuality of the                            
exchanged gluon is dominated by its transverse momentum $|k^{\prime 2}| \simeq                            
k_T^{\prime 2}$.  For clarity we have restored the subscript $T$ in this equation.                                          
                                          
If condition (\ref{eq:a3}) is imposed on the BFKL equation it can be still solved                                         
analytically.  The result is an all-order effect, which at NLO gives the large                                         
modification                                        
\be                                        
\label{eq:b3}                                        
\lambda \; \simeq \; \overline{\alpha}_S \: 4 \ln 2 (1 - 4.2 \overline{\alpha}_S)                                        
\ee                                        
of the LO value.  However it is found that the all-order correction is a much milder                                         
modification, although still significant.  A related result can be found in                                         
ref.~\cite{SALAM}.  We can therefore make the BFKL equation (\ref{eq:a1}) for the                                         
gluon much more realistic by imposing the consistency condition (\ref{eq:a3}), as                                         
well as by allowing the coupling $\alpha_S$ to run.                                        
                                        
Moreover we can extend its validity to cover the full range of $x$.  To do this we note                                         
that the BFKL equation embodies the important double leading $\log$ part of DGLAP                                         
evolution which is driven just by the singular $1/z$ part of the splitting function                                         
$P_{gg}$.  To obtain a reliable description throughout the full $x$ range (and not just                                         
at small $x$) we must include the remaining terms in $P_{gg}$, together with the                                         
quark to gluon transitions. We also introduce in eq. \ref{eq:a1} the parameter 
$k_o^2$ ($k_o^2 \simeq	1 GeV^2$) which divides the non-perturbative
($k'^2 < k_o^2$) from the perturbative ($k'^2>k_o^2$) region.
  Finally we note that the contribution from the infrared                                         
region $k^{\prime 2} < k_0^2$ in (\ref{eq:a1}) may be expressed \cite{KMS} in                                         
terms of the integrated gluon distribution at scale $k_0^2$, that is $g (x, k_0^2)$.  All                                         
the above modifications of (\ref{eq:a1}) are encapsulated in a unified BFKL/DGLAP                                         
equation of the form                                        
\newpage             
\bea                                            
\label{eq:a4}                                            
& & f(x, k^2) \; = \; \tilde{f}^{(0)} (x, k^2) + \nonumber \\                                           
& & \nonumber \\               
& & + \: \overline{\alpha}_S (k^2) k^2 \int_x^1 \frac{dz}{z}                                             
\int_{k_0^2} \frac{dk^{\prime 2}}{k^{\prime 2}} \left\{\frac{f                                            
\left( {\displaystyle \frac{x}{z}}, k^{\prime 2} \right) \Theta \left({\displaystyle                                           
\frac{k^2}{z}} - k^{\prime 2}\right) - f \left({\displaystyle \frac{x}{z}}, k^2\right)}                                            
{| k^{\prime 2} - k^2 |} \; + \; \frac{f \left({\displaystyle \frac{x}{z}}, k^2                                           
\right)}{[4k^{\prime 4}+ k^4]^{\frac{1}{2}}} \right \} \\                                            
& & \nonumber \\                                            
& & + \: \overline{\alpha}_S (k^2) \int_x^1 \frac{dz}{z} \left(\frac{z}{6}                                            
P_{gg} (z) - 1 \right ) \int_{k_0^2}^{k^2}  \frac{dk^{\prime 2}}                                            
{k^{\prime 2}} f \left(\frac{x}{z}, k^{\prime 2} \right ) \: + \:                                            
\frac{\alpha_S (k^2)}{2\pi} \int_x^1 dz P_{gq} (z) \Sigma                                             
\left(\frac{x}{z}, k^2 \right ). \nonumber                                         
\eea                                            
Now the driving term has the form                                            
\be                                          
\label{eq:a5}                                          
\tilde{f}^{(0)} (x, k^2) \; = \;  f^{(0)} (x, k^2) \: + \:                                            
\frac{\alpha_S (k^2)}{2\pi} \int_x^1 dz P_{gg} (z) \frac{x}{z} g                                            
\left(\frac{x}{z}, k_0^2 \right)                                          
\ee                                          
where in the perturbative domain $k^2 > k_0^2$ we may safely neglect the genuinely                                           
non-perturbative contribution $f^{(0)} (x, k^2)$ which is expected to decrease rapidly                                           
with increasing $k^2$.  It is important to note that (\ref{eq:a4}) only involves $f (x,                                           
k^2)$ in the perturbative region $k^2 > k_0^2$.  The input (\ref{eq:a5}) is provided                                           
by the conventional gluon at scale $k_0^2$, just as in pure DGLAP evolution.                                         
                                         
The last term in (\ref{eq:a4}) is the contribution of the singlet quark distribution to                                          
the gluon, with                                         
\be                                         
\label{eq:a6}                                         
\Sigma \; = \; \sum_q \: x (q + \bar{q}) \; = \; \sum_q (S_q + V_q)                                         
\ee                                          
where $S$ and $V$ denote the sea and valence quark momentum distributions.  The                                          
gluon, in turn, helps to drive the sea quark distribution through the $g \rightarrow                                          
q\bar{q}$ transition.  Thus equation (\ref{eq:a4}) has to be solved simultaneously                                          
with an equivalent equation for $\Sigma (x, k^2)$.                                         
                                         
Consider for the moment just the $g \rightarrow q\bar{q}$ contribution to $S_q$.                                           
The $k_T$ factorization theorem gives \cite{KTFAC}                                        
\be                                         
\label{eq:a7}                                         
S_q (x, Q^2) \; = \; \int_x^1 \: \frac{dz}{z} \: \int  \: \frac{dk^2}{k^2} \: S_{\rm                                          
box}^q \: (z, k^2, Q^2) \: f \left  ( \frac{x}{z}, k^2 \right )                                        
\ee                                          
where $S^{\rm box}$ describes the quark box (and crossed-box) contribution shown                                          
in Fig.~1.  $S^{\rm box}$ implicitly includes an integration over the transverse                                          
momentum $\kappa$ of the exchanged quark.  We have                                         
\bea                                         
\label{eq:a8}                                         
S_q^{\rm box} (z, k^2, Q^2) & = & \frac{Q^2}{4 \pi^2 k^2} \int_0^1 \: d\beta \: \int \:                                          
d^2 \kappa^\prime \alpha_S \left \{ [\beta^2 + (1 - \beta)^2 ] \:  \left (                                          
\frac{\mbox{\boldmath $\kappa$}}{D_{1 q}} \: - \:  \frac{\mbox{\boldmath                                          
$\kappa$} - \mbox{\boldmath $k$}}{D_{2q}}  \right )^2 \right . \nonumber \\                                            
& & \nonumber \\                                            
& + & [m_q^2 \: + \: 4Q^2 \beta^2 (1 - \beta)^2 ] \:  \left (\frac{1}{D_{1 q}} \: - \:                                          
\left . \frac{1}{D_{2q}} \right  )^2 \right \} \: \delta (z - z_0)                                         
\eea                                         
where $\mbox{\boldmath $\kappa^\prime$} = \mbox{\boldmath $\kappa$} - (1 - \beta) \mbox{\boldmath $k$}$ and                                         
\bea                                         
\label{eq:a9}                                         
D_{1q} & = & \kappa^2 \: + \: \beta (1 - \beta) Q^2 \: + \: m_q^2                                            
\nonumber \\                                            
& & \nonumber \\                                            
D_{2q} & = & (\mbox{\boldmath $\kappa$} - \mbox{\boldmath                                            
$k$})^2 \: + \: \beta (1 - \beta) Q^2 \: + \: m_q^2                                            
\nonumber \\                                            
& & \nonumber \\                                            
z_0 & = & \left [ 1 \: + \: \frac{\kappa^{\prime 2} +  m_q^2}{\beta (1 - \beta) Q^2} \:                                         
+  \: \frac{k^2}{Q^2} \right ]^{- 1}.                                             
\eea                                            
Eqs.~(\ref{eq:a6})--(\ref{eq:a8}) enable us to evaluate the singlet quark distribution                                         
$\Sigma$ in terms of the gluon $f$. We obtain                                        
\bea                                          
\label{eq:a10}                                          
\Sigma (x, k^2) & = & S_{\rm non-p} (x) \: + \: \sum_q \int_{x}^a                                          
\frac{dz}{z} \: S_q^{\rm box} (z, k^{\prime 2} = 0, k^2) \frac{x}{z} \: g                                          
\left(\frac{x}{z},  k_0^2 \right) \nonumber \\                                          
& & \nonumber \\                                          
& & + \: \sum_q \int_{k_0^2}^\infty \frac{dk^{\prime                                          
2}}{k^{\prime 2}} \int_x^1 \frac{dz}{z} \: S_q^{\rm box} (z, k^{\prime 2},                                          
k^2) f \left (\frac{x}{z}, k^{\prime 2} \right) \: + \: V (x, k^2)  \nonumber  \\                                          
& & \\                                          
& & + \: \int_{k_0^2}^{k^2} \frac{dk^{\prime 2}}{k^{\prime 2}} \:                                          
\frac{\alpha_S  (k^{\prime 2})}{2\pi} \int_x^1 dz \: P_{qq} (z)                                          
S_{uds} \left(\frac{x}{z}, k^{\prime 2} \right ) \nonumber                                          
\eea                                         
where $a = (1 + 4m_q^2/Q^2)^{-1}$ and $V = x (u_v + d_v)$.  Here we have                                    
separated off the non-perturbative contributions.  $S_{\rm non-p}$ is the contribution                                    
from the region $k^2, \kappa^{\prime 2} < k_0^2$ and the next term is the                                    
contribution from the region $k^2 < k_0^2 < \kappa^{\prime 2}$.  The details are                                    
explained in ref.~\cite{KMS}.  An $S \rightarrow S$ contribution (from the light $u,                                    
d, s$ quarks) is also included.  For the light $u, d, s$ quarks $S_q^{\rm box} (z,                                    
k^{\prime 2} = 0, k^2)$ in (\ref{eq:a10}) is defined with the $\kappa^{\prime 2}$                                    
integration restricted to the region $\kappa^{\prime 2} > k_0^2$.                                
                                
In this way coupled integral equations are obtained for the unintegrated gluon $f (x,                                 
k^2)$ and the integrated quark singlet $\Sigma (x, k^2)$ distributions.  The driving                                 
terms are specified by an economically parametrized non-perturbative contribution                                 
$S_{\rm non-p} (x)$ and by the integrated gluon distribution at scale $k_0^2 =                                 
1$~GeV$^2$ which was taken to be of the form                                
\be                                
\label{eq:b10}                                
xg (x, k_0^2) \; = \; N (1 - x)^\beta.                                
\ee                                
The valence distribution $V (x, k^2)$ was taken from the parton set of                                 
ref.~\cite{GRV}.  With this input the coupled equations (\ref{eq:a4}) and                                 
(\ref{eq:a10}) were solved and a fit made of the deep inelastic electron-proton $F_2$                                 
data obtained by the H1, ZEUS, NMC and BCDMS collaborations \cite{KMS}.  An                                 
excellent description of these data was obtained with physically reasonable values of                                 
the parameters:  $N = 1.57$ and $\beta = 2.5$.  Incidentally, the output gluon at $x                                 
\sim 0.4$ was reasonably compatible with the expectations of prompt photon data.  In                                 
summary, ref.~\cite{KMS} gives as unintegrated gluon distribution and, through                                 
$k_T$ factorization, a sea distribution which can be reliably extrapolated to very low                                 
values of $x$. \\                                
                               
\newpage             
\noindent {\large \bf 3.~~The neutrino cross sections}                                        
                                        
Here we collect together all the relevant formulas which are needed to calculate the                                         
cross sections for neutrino-nucleon interactions.  We express the cross sections in                                         
terms of the structure functions for an isoscalar nucleon target, $N = (n + p) / 2$,                                         
\cite{DR,HM,ESW}                                        
\bea                                        
\label{eq:a11}                                        
\frac{d^2 \sigma^{\nu, \overline{\nu}}}{d x d y} & = & \frac{G_F ME}{\pi} \left (                                         
\frac{M_i^2}{Q^2 + M_i^2} \right )^2 \: \left \{ \frac{1 + (1 - y)^2}{2} \: F_2^\nu (x,                                         
Q^2) \right . \nonumber \\                                        
& & \\                                        
& & - \: \left . \frac{y^2}{2} \: F_L^\nu (x, Q^2) \: \underline{+} \: y \left ( 1 -                                         
\frac{y}{2} \right ) \: x F_3^\nu (x, Q^2) \right \} \nonumber                                        
\eea                                        
where $G_F$ is the Fermi coupling constant, $M$ is the proton mass, $E$ is the                                         
laboratory energy of the neutrino and $y = Q^2/xs$.  The mass $M_i$ is either                                         
$M_W$ or $M_Z$ according to whether we are calculating charged current (CC) or                                         
neutral current (NC) neutrino interactions.                                        
                                        
We may express the structure functions $F_i$ in terms of the valence and sea quark                                         
momentum distributions, $V_q$ and $S_q$, of (\ref{eq:a8}).  For the charged current                                         
$\nu N$ interaction we have                                        
\be                                   
\label{eq:a12}                                        
F_2^{\rm CC} \; = \; x (u_v + d_v) \: + \: S_u \: + \: S_d + 2x (s + c + b + t),                                   
\ee                                   
where the heavy quark contributions are calculated from the photon-gluon fusion                                         
mechanism.  To be precise for the $s \rightarrow c$ and $\bar{c} \rightarrow \bar{s}$                                         
contributions we use                                        
\bea                                        
\label{eq:a13}                                        
2 x q (x, Q^2) & = & \int_x^{a_c (k^2 = 0)} \: \frac{dz}{z} \: S_q^{\rm box} (z, k^2                                    
= 0, Q^2) \: \frac{x}{z} \: g \left ( \frac{x}{z}, k_0^2 \right ) \nonumber \\                                        
& & \\                                        
& & + \: \int_{k_0^2}^{\infty} \: \frac{dk^2}{k^2} \: \int_x^{a_c (k^2)} \:                                    
\frac{dz}{z} \: S_q^{\rm box} (z, k^2, Q^2) \: f \left ( \frac{x}{z}, k^2 \right )                                    
\nonumber                                        
\eea                                        
with $q = s$ or $c$ and with $S_q^{\rm box}$ defined by (\ref{eq:a8}) with $m_q =                                         
0$.  However the mass is included in the threshold factor                                        
\be                                        
\label{eq:a14}                                        
a_c (k^2) \; = \; \left [ 1 \: + \: \frac{k^2 + m_c^2}{Q^2} \right ]^{-1}.                                        
\ee                                        
For the small $b \rightarrow t$ and $\bar{t} \rightarrow \bar{b}$ contributions we use                                         
the standard on-shell factorization formula                                        
\be                                        
\label{eq:a15}                                        
2xq (x, Q^2) \; = \; \int_x^a \: \frac{dz}{z} \: H (z, m_q, m_{q^\prime}, Q^2) \:                                         
\frac{x}{z} \: g \left ( \frac{x}{z}, \hat{s} \right )                                        
\ee                                        
with $q = b$ or $t$ (and $q^\prime = t$ or $b$) and where the scale $\hat{s} = Q^2 (1                                    
- z)/z$ and                                
\be                                
\label{eq:z15}                                
a \; = \; [1 + (m_t + m_b)^2/Q^2]^{-1}.                                
\ee                                
The functions $H$ are defined in ref.~\cite{GGR}                                   
\bea                                   
\label{eq:b15}                                   
F_3^{\rm CC} & = & u_v \: + \: d_v \nonumber \\                                        
& & \\                                        
F_L^{\rm CC} & = & B_u \: + \: B_d \: + \: \frac{4 \alpha_S (Q^2)}{3 \pi} \: \int_x^1                                    
\: dy \left ( \frac{x}{y} \right )^2 \: F_2^{\rm CC} (y, Q^2). \nonumber                                   
\eea                                   
The function $B_q$ describes the boson-gluon fusion contribution to $F_L$                                        
\bea                                        
\label{eq:c16}                                        
B_q (x, Q^2) & = & \frac{\alpha_S (Q^2)}{\pi} \: \int_x^1 \: \frac{dy}{y} \:                                         
\frac{x}{y} \left ( 1 - \frac{x}{y} \right ) \: y g (y, k_0^2) \nonumber \\                                        
& & \\                                        
& & + \: \frac{Q^4}{\pi^2} \: \int_{k_0^2} \: \frac{dk^2}{k^4} \: \int_0^1 \: d \beta                                         
\: \beta^2 (1 - \beta)^2 \: \int \: d^2 \kappa^\prime \left ( \frac{1}{D_{1q}} -                                         
\frac{1}{D_{2q}} \right )^2 \: f \left ( \frac{x}{z_0}, k^2 \right ) \nonumber                                        
\eea                                        
where $D_{iq}$ and $z_0$ are given in (\ref{eq:a9}).  Besides $B_u$ and $B_d$,                                    
there is also boson-gluon production of heavy quarks which may be treated similarly                                    
to that for $F_2^{\rm CC}$.                                   
                                        
The structure functions for the neutral current $\nu N$ interaction are                                        
\be                                   
\label{eq:a16}                                        
F_2^{\rm NC} \; = \; (L_u^2 + L_d^2 + R_u^2 + R_d^2) \: \frac{1}{4} \left                                         
\{\sum_q S_q \: + \: x u_v \: + \: x d_v \right \},                                 
\ee                                   
where the sum over $S_q$ is given by the first three terms on the right hand side of                                    
(\ref{eq:a10}), and                                   
\bea                                   
\label{eq:b16}                                        
F_3^{\rm NC} & = & (L_u^2 + L_d^2 - R_u^2 - R_d^2) \: \frac{1}{4} (u_v + d_v)                                    
\nonumber \\                                        
& & \\                                        
F_L^{\rm NC} & = & (L_u^2 + L_d^2 + R_u^2 + R_d^2) \: \frac{1}{4} \sum_q \:                                         
B_q \: + \: \frac{4 \alpha_S (Q^2)}{3 \pi} \: \int_x^1 \: dy \left ( \frac{x}{y} \right )^2                                         
\: F_2^{\rm NC} (y, Q^2) \nonumber                                       
\eea                                        
where the chiral couplings are                                        
\bea                                        
\label{eq:a17}                                        
L_u \; = \; 1 \: - \: \textstyle{\frac{4}{3}} \: \sin^2 \theta_W, & & L_d \; = \; -1 \: + \:                                    
\textstyle{\frac{2}{3}} \: \sin^2 \theta_W, \nonumber \\                                        
& & \\                                        
R_u \; = \; - \textstyle{\frac{4}{3}} \: \sin^2 \theta_W, & & R_d \; = \;                                    
\textstyle{\frac{2}{3}} \: \sin^2 \theta_W. \nonumber                                        
\eea                                       
                                
The parton distributions that are used to evaluate the neutrino cross sections are those                                 
described in Section 2, which are obtained by solving the unified BFKL/DGLAP                                 
equations with higher order $\ln 1/x$ effects incorporated via the consistency                                 
condition.  The various components of the CC and NC neutrino cross sections are                                 
shown as a function of the laboratory neutrino energy in Figs.~2 and 3 respectively.                                  
In each case we see that for neutrino energies $E > 10^5$~GeV the sea quark                                 
contributions dominate over the valence.  The rise of the sea distributions with                                 
decreasing $x$ is reflected in the continued rise of the sea components of the cross                                 
section with energy.  We also note that for $E > 10^5$~GeV the valence component                                 
of the cross section becomes independent of energy.  This results from the fact that                                 
when (\ref{eq:a11}) is integrated over $x$ and $y$, or to be precise over $x$ and                                 
$Q^2$, the $Q^2$ integration is effectively cut-off at $Q^2 \sim M_i^2$, together                                 
with the fact that the number of valence quarks is finite.  The threshold effects of the                                 
heavy quark contributions are also evident in Figs.~2 and 3.  We combine the charged                            
current and neutral current cross sections in Fig.~4.                           
                                
In ref.~\cite{KMS} we also solved the unified equation with the omission of                                 
non-leading effects arising from the consistency condition.  Using those partons we                                 
obtain higher cross sections at ultrahigh energies as illustrated by the dashed curve in                                 
Fig.~5.  Of course this is only shown for comparison since it is based on LO BFKL                            
which is an unreliable approximation \cite{NLO}. Therefore throughout this paper we                            
impose the consistency constraint (\ref{eq:a3}) which generates the dominant non-leading $\ln(1/x)$ effects.  In Fig.~6 a comparison is made with                            
other recent calculations of the $\nu N$ total cross section \cite{GQRS1,GKR} based  on NLO DGLAP evolution, with BFKL and higher order $\ln (1/x)$ effects                            
neglected. We see that our results and those of \cite{GQRS1,GKR} are remarkably  
similar considering their different dynamic origin.  Since we include resummation of  
$\ln (1/x)$ effects we expect the continuous curve to give the most reliable  
extrapolation to ultrahigh energies.  In fact we may conclude from a comparison of  
(\ref{eq:b2}) and (\ref{eq:b3}) that our all-order sub-leading $\ln (1/x)$ treatment  
gives an upper estimate of the cross sections.  Considering that the predictions are so  
similar, although they are based on different dynamical assumptions, we may  
conclude that the ambiguity in extrapolating the neutrino cross sections to ultrahigh  
energies is less than might at first be expected.                           
                           
At ultrahigh energies the antineutrino cross sections are essentially identical to the                            
neutrino cross sections, since the difference is due to the structure function $F_3$                            
which is controlled by valence quarks.  In Table I we list the various cross sections as                            
a function of energy.                           
                           
To gain insight into the $(x, Q^2)$ domain that is sampled by ultrahigh energy                            
neutrinos we show plots in Fig.~7 in which the contours enclose regions contributing                            
different fractions of the total cross section.  The two plots illustrate the dependence                            
of the domains on the neutrino energy.  For the ultrahigh energy chosen for Fig.~7(b)                            
we see that the main contribution comes from the domain $Q^2 \simeq M_W^2$ and                            
$x \sim M_W^2/(2M E)$, where $M$ is the proton mass, as anticipated from                            
(\ref{eq:a11}).  For the lower energy used in Fig.~7(a) some residual propagator                            
effects are still apparent and $Q^2 \lapproxeq M_W^2$ and $x \lapproxeq                            
M_W^2/2M E$. \\                                 
                                   
\noindent {\large \bf 4.~~Transport equations}                                   
                                   
The ultrahigh energy neutrinos when penetrating through the Earth can undergo                                    
attenuation due to charged and neutral current interactions as well as the                                    
regeneration  due to the neutral current interactions at higher energies.                                     
Both effects are summarized in the transport equation for the neutrino flux                                    
$I(E,\tau)$ \cite{NIK,NP}:                                    
\be                                   
\label{eq:a18}                                   
{dI(E,\tau)\over d\tau} \; = \; - \sigma_{\rm TOT}(E) I(E,\tau) \: + \: \int {dy\over 1-y}                                    
{d\sigma_{\rm NC}(E^{\prime},y) \over dy} I(E^{\prime},\tau)                                   
\ee                                   
where $\sigma_{\rm TOT} = \sigma_{\rm CC} + \sigma_{\rm NC}$ and where $y$                                    
is, as usual, the fractional energy loss such that                                   
\be                                   
\label{eq:a19}                                   
E^{\prime}={E\over 1-y}.                                   
\ee                                  
The variable $\tau$ is the number density of nucleons $n$ integrated along a path of                                    
length $z$ through the Earth                                   
\be                                   
\label{eq:a20}                                   
\tau \; = \; \int_0^z \: dz^\prime \: n (z^\prime).                                   
\ee                 
The number density $n(z)$ is defined as $n(z)=N_A \rho (z) $ where $\rho (z)$ is the density of Earth along the neutrino path length $z$ and $N_A$ is the Avogadro number.                  
Clearly the number of nucleons $\tau$ encountered along the path $z$ depends upon                                    
the nadir angle $\theta$ between the normal to the Earth's surface (passing through the                                    
detector) and the direction of the neutrino beam incident on the detector.  For example                                    
$\theta = 0^\circ$ corresponds to a beam transversing the diameter of the Earth.                                     
To compute the variation of $\tau$ with the angle $\theta$ we need to know the                                    
density profile of the Earth.  We use the preliminary Earth model \cite{EARTH}.                                   
                                   
It is convenient to represent the solution of the transport                                    
equation (\ref{eq:a18})                                    
in the form:                                    
\be                                   
\label{eq:a21}                                   
I(E,\tau)=I_0(E) \exp (- \sigma_{\rm TOT} (E) \tau) \: \Psi(E,\tau)                                   
\ee                                   
where $I_{0}(E)$ denotes the flux of neutrinos incident on the surface of the                                    
Earth from outer space.  The function $\Psi(E,\tau)$ would be unity in the absence of                                    
regeneration.  It satisfies the following equation:                                    
\be                                  
\label{eq:a22}                                   
{ d\Psi(E,\tau)\over d\tau} \; = \; \int {dy\over 1-y} {I_0(E^{\prime})\over I_0(E)}                                  
\exp[-(\sigma_{\rm TOT}(E^{\prime}) - \sigma_{\rm TOT}(E))\tau]{d\sigma_{\rm                                  
NC}(E^{\prime},y) \over dy} \Psi(E^{\prime},\tau),                                   
\ee                                 
with the initial condition  $\Psi(z,\tau = 0) = 1$.
We solve this equation numerically and determine the regeneration factor $\Psi$ as a                               
function of $E$ and $\tau$.  The solution is sensitive to the behaviour of                               
$\sigma_{\rm TOT}$ for $E^\prime > E$ and on energy dependence of the initial flux                               
$I_0 (E)$ and on the value of $d \sigma_{\rm NC}/dy$.  In particular the flatter the                               
initial spectrum $I_0 (E)$ the more it is possible to sample $d \sigma/dy$ at energies                               
$E^\prime$ much higher than $E$ and so the amount of regeneration is increased.  In                            
practice the result is a combined effect of the fall-off due to $I_0$ and the                            
experimental attenuation and the enhancement due to $d \sigma_{\rm NC}/dy$.  To                            
illustrate the general properties of the solution we show the shadowing factor,                           
\be                              
\label{eq:a23}                              
S \; = \; \Psi (E, \tau) \: \exp (- \sigma_{\rm TOT} (E) \tau),                           
\ee                              
of (\ref{eq:a21}) for two different, but physically relevant, forms of the initial flux.                                
First we consider the flux of atmospheric neutrinos given by \cite{ATMOS}                           
\be                              
\label{eq:b23}                              
I_0 (E) \; = \; c E^{-3.6},               
\ee                              
which has a rapid fall-off with energy, and second, we consider the flux from Active  
Galactic Nuclei (AGN) as given by ref.~\cite{AGN1} for which the incident flux $I_0 
(E)$ is approximately constant throughout the interval $10^3 < E < 10^5$~GeV.  The 
results are presented by the continuous curves in Fig.~8 for three different incident 
directions, $\theta = 0^\circ,                            
40^\circ$ and $80^\circ$, corresponding  to values of $\tau / N_A = 1.1 \times 10^{10}, 0.45                   
\times 10^{10}$, and $0.072 \times 10^{10}$~cmwe respectively.  For illustration we                   
also show, by dashed curves, the pure attenuation factors $A \equiv \exp (-  
\sigma_{\rm                   
TOT} \tau)$ with regeneration omitted.  Since the neutrino cross sections increase                   
with energy we observe that the attenuation factor                            
$A$ leads to total shadowing once the energy is sufficiently high.  The energy at                            
which this occurs depends mainly on $\tau$.  The regeneration, which increases the                            
flux at the detector, is sensitive to the energy dependence of $I_0$.  For a steeply                            
falling flux, corresponding, for example, to atmospheric neutrinos, regeneration gives                            
a rather small effect.  On the other hand for a flatter initial flux $I_0 (E)$ the                 
regeneration effect                            
can be quite significant.  In fact it can even enhance the initial flux by as much as                            
40\%, as can be seen from the shadowing factor at $\theta = 0^\circ$ for the AGN flux                            
used in Fig.~6.  A similar result was found in ref.~\cite{NIK}.  In the next section we                            
present the flux arriving at the detector from various sources for a range of angles                            
taking into account the full shadowing factor of (\ref{eq:a21}). \\                             
                              
\noindent {\large \bf 5.~~Penetration of the Earth for given incident fluxes}                                 
                                
The initial incident neutrino flux $I_0 (E)$ is modified on its passage through the                            
Earth by the shadowing factor $S$ of (\ref{eq:a23}).  For experimental purposes the                            
relevant quantity is the neutrino flux $I (E)$ reaching the detector at different nadir                            
angles $\theta$. The four plots in each of Figs.~9 and 10 show first the initial flux  
$I_0 (E)$ and then the flux at the detector $I (E)$ for the three nadir angles $\theta =  
80^\circ, 40^\circ$ and $0^\circ$.  Recall that $0^\circ$ is penetration of neutrinos  
through the centre of the Earth.  The various curves are for neutrinos of different  
origin.  For convenience of reference we show the atmospheric neutrino                            
flux in both Figs.~9 and 10.  In Fig.~9 we compare this background spectrum                            
with three different models of the flux expected from Active Galactic Nuclei (AGN)                            
\cite{AGN1,AGN2,AGN3}.  The AGN flux stands out above the atmospheric                            
neutrino background for neutrino energies above about $10^5$~GeV.  However the                            
AGN spectrum is attenuated at ultrahigh neutrino energies by shadowing.  The smaller                            
the nadir angle $\theta$ the greater the shadowing.  For example at $\theta = 80$ or                            
$40^\circ$ the weighted flux $E dN/dE$ falls below                            
$10^{-15}$~cm$^{-2}$~s$^{-1}$~sr$^{-1}$ at $E \sim 10^8$ and $10^7$~GeV                 
respectively.                             
Fig.~10 shows the corresponding fluxes for neutrinos coming from gamma ray bursts                            
\cite{GRB} and from a sample top-down model \cite{TDM}.  Similar attenuation can                            
be observed in Fig.~10 to that in Fig.~9.                            
                           
So far we have discussed the effects generated by inelastic neutrino interactions with                                 
hadronic matter.  Another possible source of modification of the neutrino fluxes                                 
which penetrate the Earth may be due to neutrino oscillations                            
\cite{GALLEX,SAGE,KAM,SUPKAM}.  The properties of neutrino oscillations in                            
matter have been discussed in \cite{OSC}.  They are characterised by an effective                            
oscillation length $\ell_{\rm m}$ and mixing angle $\theta_{\rm m}$ which differ                            
from those in vacuum, which we denote $\ell_{\rm v}$ and $\theta_{\rm v}$.  To be                            
precise in the MSW model for mixing of two neutrino species we have                                
\bea                                
\label{eq:a24}                                
\ell_{\rm m} & = & \ell_{\rm v} \left [ 1 + 2 \frac{\ell_{\rm v}}{\ell} \: \cos                                 
\theta_{\rm v} + \frac{\ell_{\rm v}^2}{\ell^2} \right ]^{- \frac{1}{2}} \\                                
& & \nonumber \\                                
\label{eq:a25}                                
\tan 2 \theta_{\rm m} & = & \frac{\sin 2 \theta_{\rm v}}{\cos 2 \theta_{\rm v} +                                 
(\ell_{\rm v}/\ell)}                                
\eea                                
where $\ell$ originates from the matter contribution to the oscillation length and is                                 
given by                                
\be                                
\label{eq:a26}                                
\ell \; = \; \frac{\sqrt{2} \pi}{G_F n_{\it e}} \; = \; \frac{1.77 \times 10^7}{\rho_{\it                                 
e}} \: {\rm m}                                
\ee                               
where $n_{\it e}$ is the electron density and $\rho_{\it e}$ is the number density in                                
units of Avogadro's number/cm$^3$.  An important property of $\ell_{\rm m}$ is that                                
it saturates at the value $\ell$ given by (\ref{eq:a26}) \cite{ITALY}.  That is                               
\be                               
\label{eq:a27}                               
(\ell_{\rm m})_{\rm max} \; = \; \ell \; \simeq \; 10^3 - 10^4~{\rm km}                               
\ee                               
for matter oscillations in the Earth with $\rho_{\it e} = 2 - 10$.  This is in contrast                                
with the oscillation length in vacuum                               
\be                               
\label{eq:a28}                               
\ell_{\rm v} \; = \; 4 \pi E/\Delta {\rm m}_\nu^2                               
\ee                               
which increases with increasing energy.  Most importantly we see that at sufficiently                               
high energy $\ell_{\rm v}/\ell$ becomes very large and the effective mixing angle                               
$\theta_m$ tends to zero, see (\ref{eq:a25}).  In fact for the very small values of the                               
mass difference of the neutrinos ($\Delta {\rm m}_\nu^2 \simeq 10^{-6}$~eV$^2$                               
suggested by the data \cite{KAM}) we can safely neglect the modification of the                               
neutrino fluxes due to neutrino oscillations in their passage through the Earth for $E >                               
1$~TeV or so.  Of course neutrino oscillations can, in principle, modify the flux $I_0                               
(E)$ arriving at the Earth. \\                             
                               
\noindent {\large \bf 6.~~Conclusions}                          
                
In this paper we have extended  the recently developed \cite{KMS} unified                
BFKL/DGLAP framework to calculate the                 
total cross sections of ultrahigh neutrino interactions with nucleons.  The framework                 
incorporated non-leading $\ln (1/x)$ effects which are generated by the                 
consistency condition given in (\ref{eq:a3}).  In this way it was possible to resum to                
all orders the                 
dominant part of the non-leading $\ln(1/x)$ effects and to obtain a physically                  
acceptable description of the  structure functions at low $x$.  Indeed this unified                
BFKL/DGLAP approach gave an excellent description of the $F_2$ data from HERA                
and enables us to  extrapolate the                 
structure functions to the region of the very small values of $x$ which are probed in                 
the ultrahigh energy neutrino interactions.  We compared our predictions for the cross                 
sections with the results of the two recent calculations which were obtained within the                 
NLO DGLAP framework.  We find that all three approaches give results for                 
the neutrino cross sections within 30\% - 40\% or so. We may conclude that the potential                 
ambiguities in the extrapolation of the cross sections are much smaller than might  
have                 
been  expected.  However the present calculation, which includes a treatment of $\ln  
(1/x)$ effects at small $x$, should be the more reliable for ultrahigh energy neutrino  
interactions.  It should be stressed that  the inclusion of the dominant             
non-leading $\ln(1/x)$ effects was crucial for obtaining this result.  Extrapolation             
based on                 
the LO BFKL equation would generate cross sections which would be enhanced by a                 
factor of more than two at ultrahigh neutrino energies $E_\nu \sim 10^12 GeV$.  This LO approximation is, however, known to be unreliable.                 
                 
The calculated neutrino cross sections were next used as an input in the transport                 
equation describing the modification of the neutrino flux during the penetration of the                
Earth.                 
This equation incorporated both the attenuation effects of the neutrino \lq\lq  
beam\rq\rq~as well as the regeneration of                 
neutrinos due to neutral current interactions.  We solved the equation for a variety                 
of initial neutrino fluxes and discussed the dependence of the shadowing factor                 
upon the nadir angle $\theta$.  We found that although attenuation of the neutrino flux                
is the major effect, nevertheless enhancement due to neutrino regeneration can                
become appreciable for fluxes                  
which extend to the very high neutrino energies, like those originating from AGN                 
sources.  The calculations of the neutrino fluxes at various nadir angles show that at                
sufficiently high energies neutrinos become strongly attenuated.  The smaller the nadir             
angle the lower the energy of complete attenuation.                 
                 
To sum up we have demonstrated that a framework which incorporates QCD  
expectations at low $x$, including the BFKL effects with the resummation                  
of the non-leading $\ln(1/x)$ terms, gives neutrino cross sections which are                
compatible                 
with those obtained within the NLO DGLAP framework.  This strongly                  
limits potential ambiguities in the possible values of the cross sections                  
extrapolated  from the HERA domain to the region of $x$ and $Q^2$ which can be                  
probed in ultrahigh energy neutrino interactions. Due to large values of these                  
cross sections the attenuation effects reduce the fluxes of ultrahigh                  
energy neutrinos particularly at small nadir angles.  Nevertheless there is a window for  
the observation of AGN by km$^3$ underground detectors of the energetic decay  
muons.  We have found that the AGN flux             
exceeds the atmospheric neutrino background for neutrinos energies $E \gapproxeq             
10^5$~GeV.  Typical results are shown in Fig.~9.  These illustrate the possibility of  
observing AGN at various nadir angles by \lq\lq neutrino astronomy\rq\rq. \\                
                
\noindent {\large \bf Acknowledgements}                
                
We thank Francis Halzen and Argyris Nicolaidis for stimulating our interest in this                 
problem.  We thank them and Ron Girdler, Karl Mannheim, Floyd Stecker and
Ubi Wichoski for                 
useful discussions and correspondence. G\"unter Sigl is thanked for providing data                 
files for neutrino fluxes.  JK and AMS thank the Physics Department and Grey                 
College of the University of Durham for their warm hospitality.  JK thanks the UK                 
PPARC for a Visiting Fellowship.  This research has been supported in part by the                 
Polish State Committee for Scientific Research (KBN) grants N0~2~P03B~89~13,                 
2~P03B~137~14 and by the EU Fourth Framework Programme \lq\lq Training and                 
Mobility of Researchers", Network \lq\lq Quantum Chromodynamics and the Deep                 
Structure of Elementary Particles", contract FMRX - CT98 - 0194.

\newpage           
\begin{table}[h]           
\caption{The charged-current and neutral-current cross sections (in cm$^2$) for $\nu             
N$ and $\overline{\nu} N$ interactions, where $N = \frac{1}{2} (p + n)$.}            
\begin{center}            
\begin{tabular}{|c|cc|cc|} \hline           
$E_\nu$~GeV & $\sigma_{\rm CC} (\nu N)$ & $\sigma_{\rm NC} (\nu N)$ &             
$\sigma_{\rm CC} (\overline{\nu} N)$ & $\sigma_{\rm NC} (\overline{\nu} N)$  \\    
\hline           
& & & & \\       
$10^1$ & $5.86 \times 10^{-38}$ & $1.81 \times 10^{-38}$ & $2.00 \times             
10^{-38}$ & $7.54 \times 10^{-39}$ \\      
& & & & \\       
$10^2$ & $6.41 \times 10^{-37}$ & $1.99 \times 10^{-37}$ & $2.99 \times    
10^{-37}$ & $1.05 \times 10^{-37}$ \\      
& & & & \\       
$10^3$ & $6.12 \times 10^{-36}$ & $1.94 \times 10^{-36}$ & $3.29 \times    
10^{-36}$ & $1.16 \times 10^{-36}$ \\           
& & & & \\      
$10^4$ & $4.59 \times 10^{-35}$ & $1.55 \times 10^{-35}$ & $3.01 \times    
10^{-35}$ & $1.07 \times 10^{-35}$  \\           
& & & & \\      
$10^5$ & $2.07 \times 10^{-34}$ & $7.33 \times 10^{-35}$ & $1.74 \times    
10^{-34}$ & $6.20 \times 10^{-35}$  \\           
& & & & \\      
$10^6$ & $6.47 \times 10^{-34}$ & $2.28 \times 10^{-34}$ & $6.19 \times    
10^{-34}$ & $2.18 \times 10^{-34}$  \\           
& & & & \\      
$10^7$ & $1.73 \times 10^{-33}$ & $5.95 \times 10^{-34}$ & $1.72 \times    
10^{-33}$ & $5.90 \times 10^{-34}$  \\           
& & & & \\      
$10^8$ & $4.33 \times 10^{-33}$ & $1.45 \times 10^{-33}$ & $4.32 \times    
10^{-33}$ & $1.45 \times 10^{-33}$  \\           
& & & & \\      
$10^9$ & $1.04 \times 10^{-32}$ & $3.38 \times 10^{-33}$ & $1.04 \times    
10^{-32}$ & $3.38 \times 10^{-33}$  \\           
& & & & \\      
$10^{10}$ & $2.40 \times 10^{-32}$ & $7.61 \times 10^{-33}$ & $2.40 \times    
10^{-32}$ & $7.61 \times 10^{-33}$  \\           
& & & & \\      
$10^{11}$ & $5.38 \times 10^{-32}$ & $1.66 \times 10^{-32}$ & $5.38 \times    
10^{-32}$ & $1.66 \times 10^{-32}$  \\           
& & & & \\      
$10^{12}$ & $1.17 \times 10^{-31}$ & $3.53 \times 10^{-32}$ & $1.17 \times    
10^{-31}$ & $3.53 \times 10^{-32}$  \\     
& & & & \\ \hline            
\end{tabular}            
\end{center}           
\end{table}            
                     
\newpage                     
\noindent {\large \bf Figure Captions}                     
                     
\begin{itemize}                     
\item[Fig.~1] Diagrammatic representation of the $k_T$-factorization formula                      
(\ref{eq:a2}).  At lowest order in $\alpha_S$ the gauge boson-gluon fusion processes,                      
$Vg \rightarrow q\bar{q}$, are given by the quark box shown (together with the                      
crossed box).  The variables $\kappa$, $k$ and $k^\prime$ denote the transverse                      
momenta of the indicated virtual particles.                     
                     
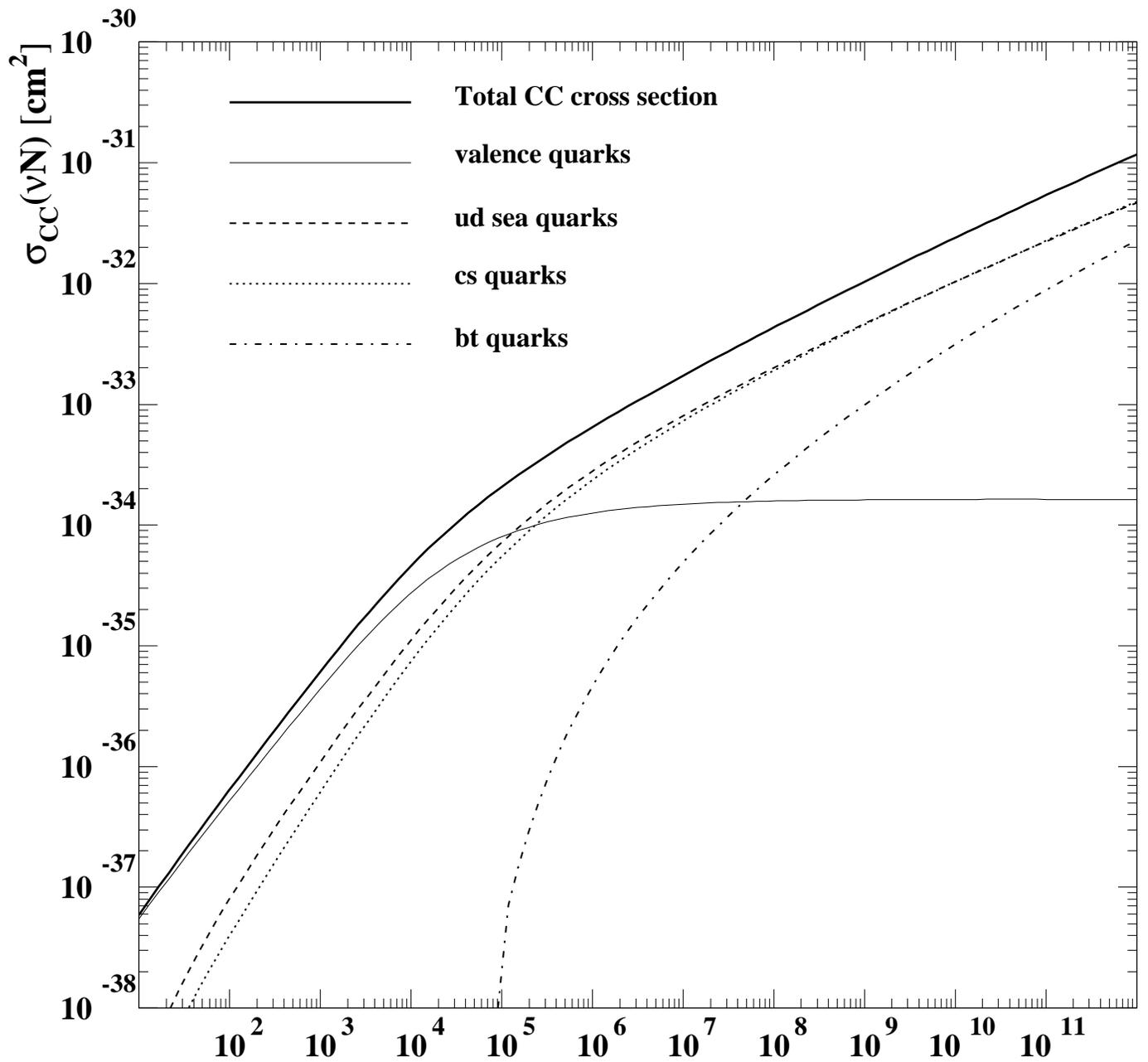
\item[Fig.~2] The total $\nu N$ charged current cross section and its decomposition                      
into components of different origin as a function of the laboratory neutrino energy.                     
                     
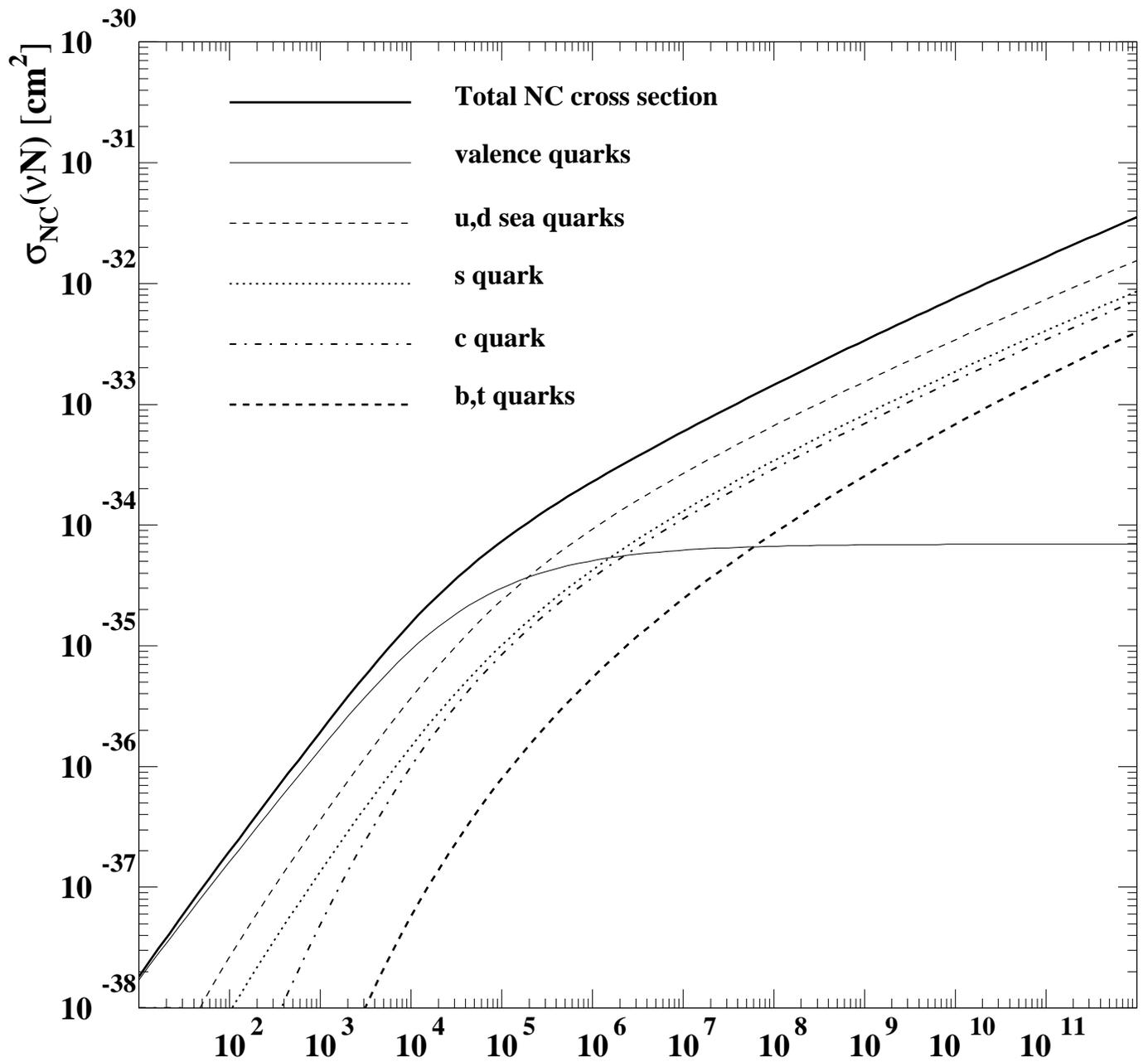
\item[Fig.~3] The total $\nu N$ neutral current cross section and its decomposition                      
into components of different origin as a function of the laboratory neutrino energy.                     
                     
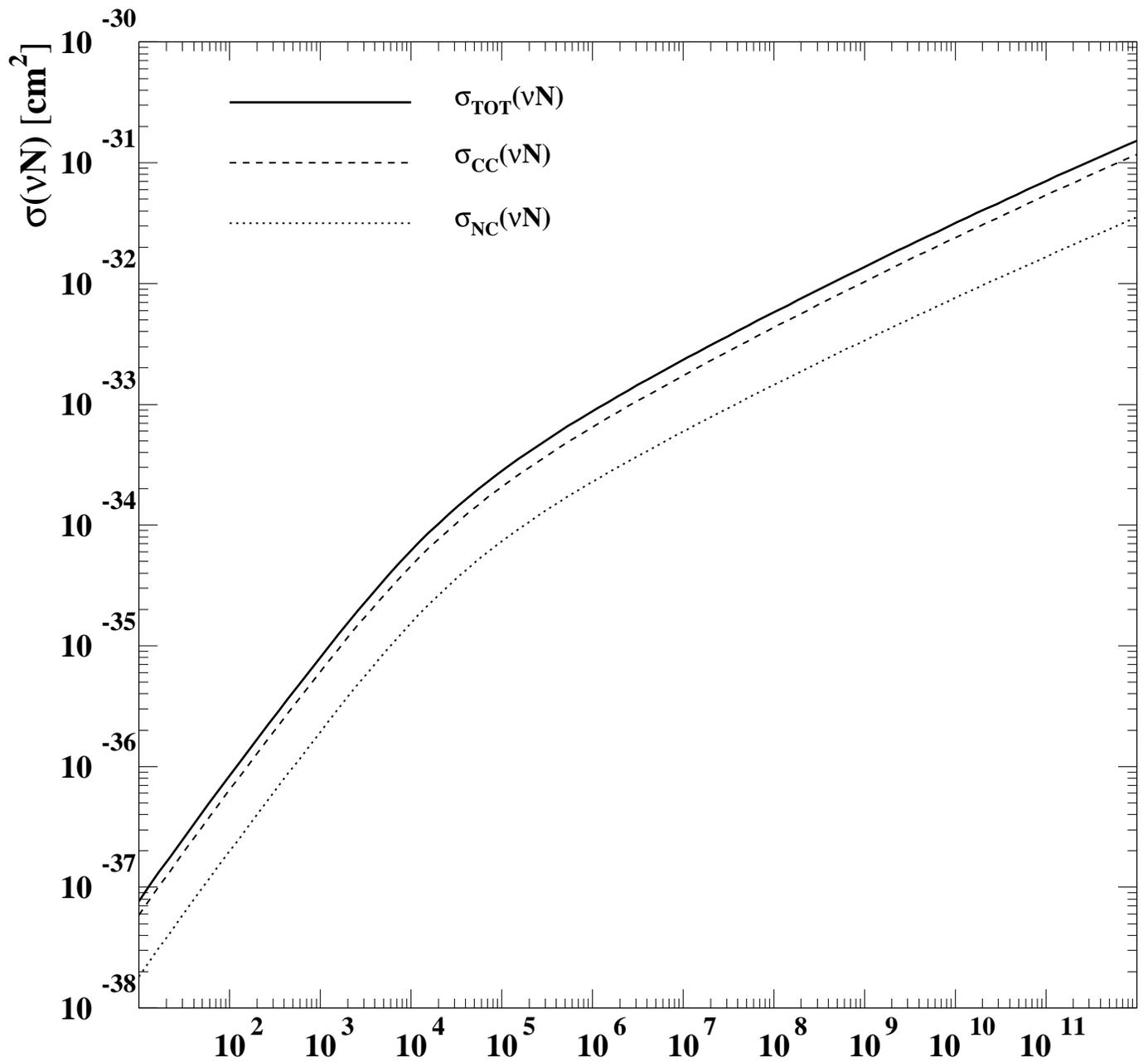
\item[Fig.~4] The total $\nu N$ cross section together with its charged current and                      
neutral current components as a function of the laboratory neutrino energy.                     
                     
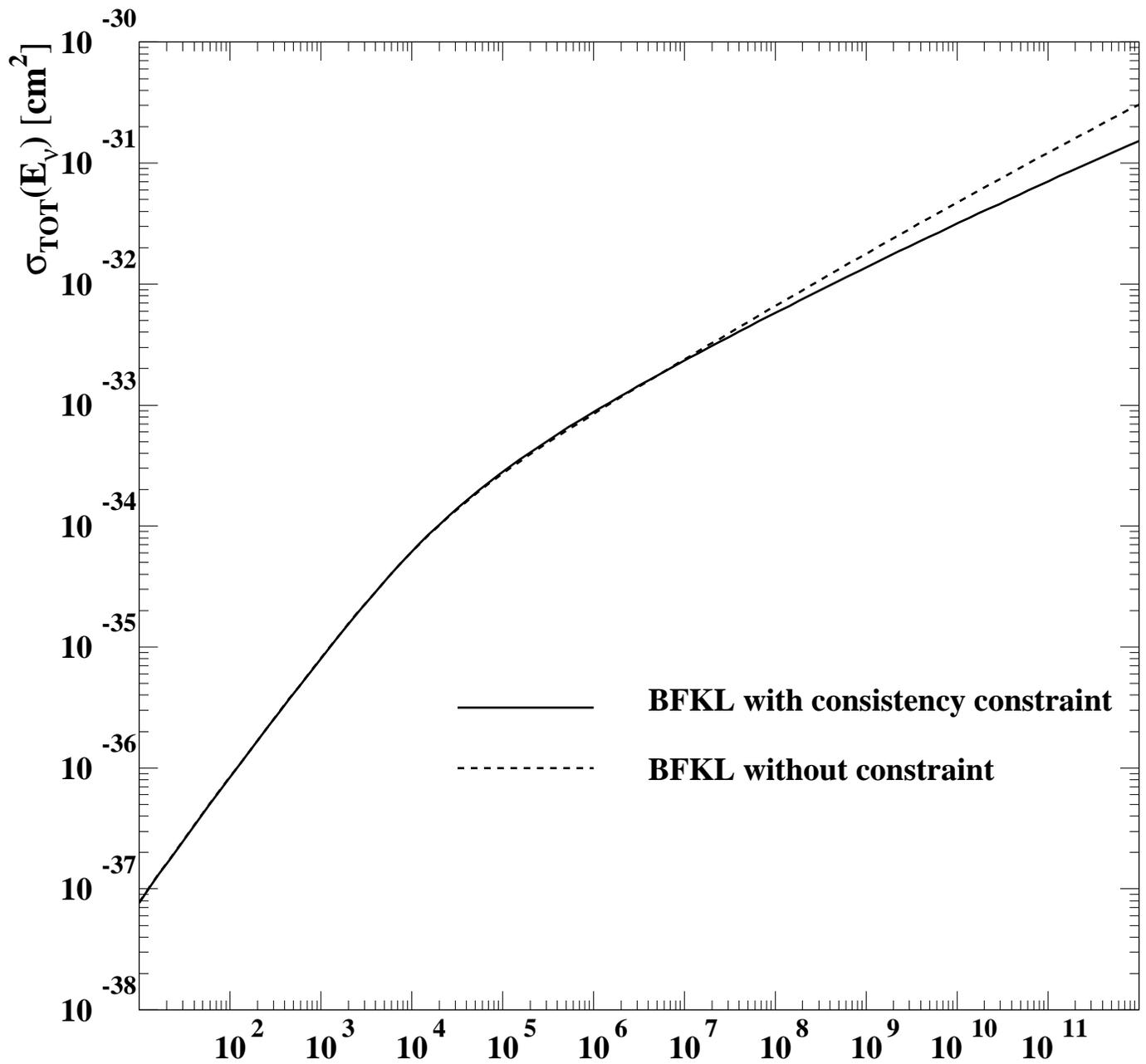
\item[Fig.~5] The comparison of the total $\nu N$ cross section calculated with and                      
without the consistency constraint (\ref{eq:a3}) imposed.                     
                     
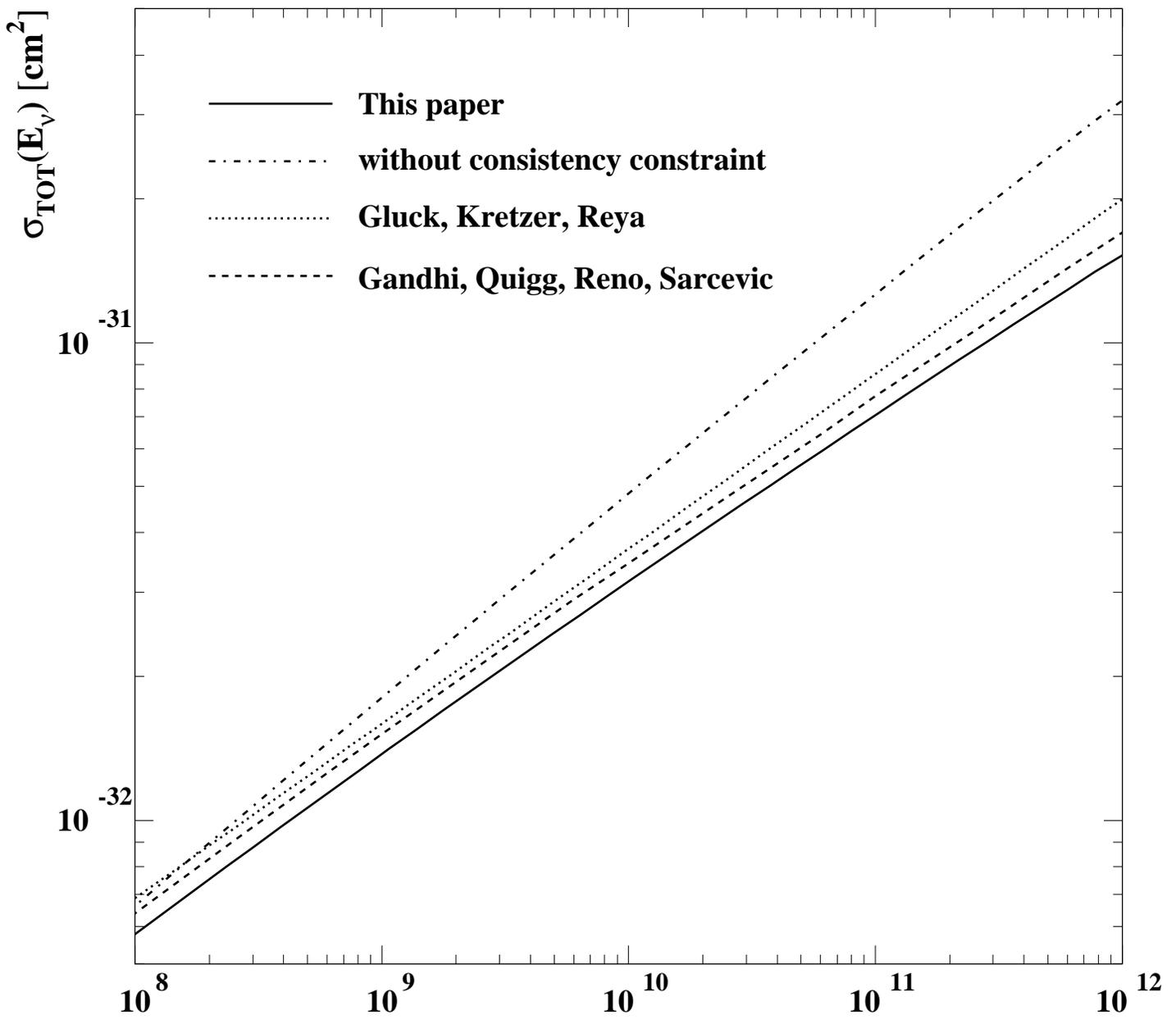
\item[Fig.~6] The prediction for the total $\nu N$ cross section obtained from a                      
unified BFKL/DGLAP equation with (and, for comparison, without) the consistency                      
condition imposed, compared to other recent calculations: \cite{GQRS1} based on
CTEQ parton distributions \cite{CTEQ} and \cite{GKR} based on GRV dynamical partons
 \cite{GRV}.                     
                     
\item[Fig.~7] A contour plot showing the $x, Q^2$ domain of the dominant                      
contribution to the \linebreak $d \sigma/d \ln (1/x) d \log Q^2$ for the total $\nu N$                     
interaction at two values of the neutrino laboratory momentum:  (a) $E_\nu =                     
10^6$~GeV and (b) $E_\nu = 10^{11}$~GeV.  The 20 contours are such that they                     
enclose a contribution of 5, 10, 15 $\cdots$ \% of the above differential cross section.                     
                     
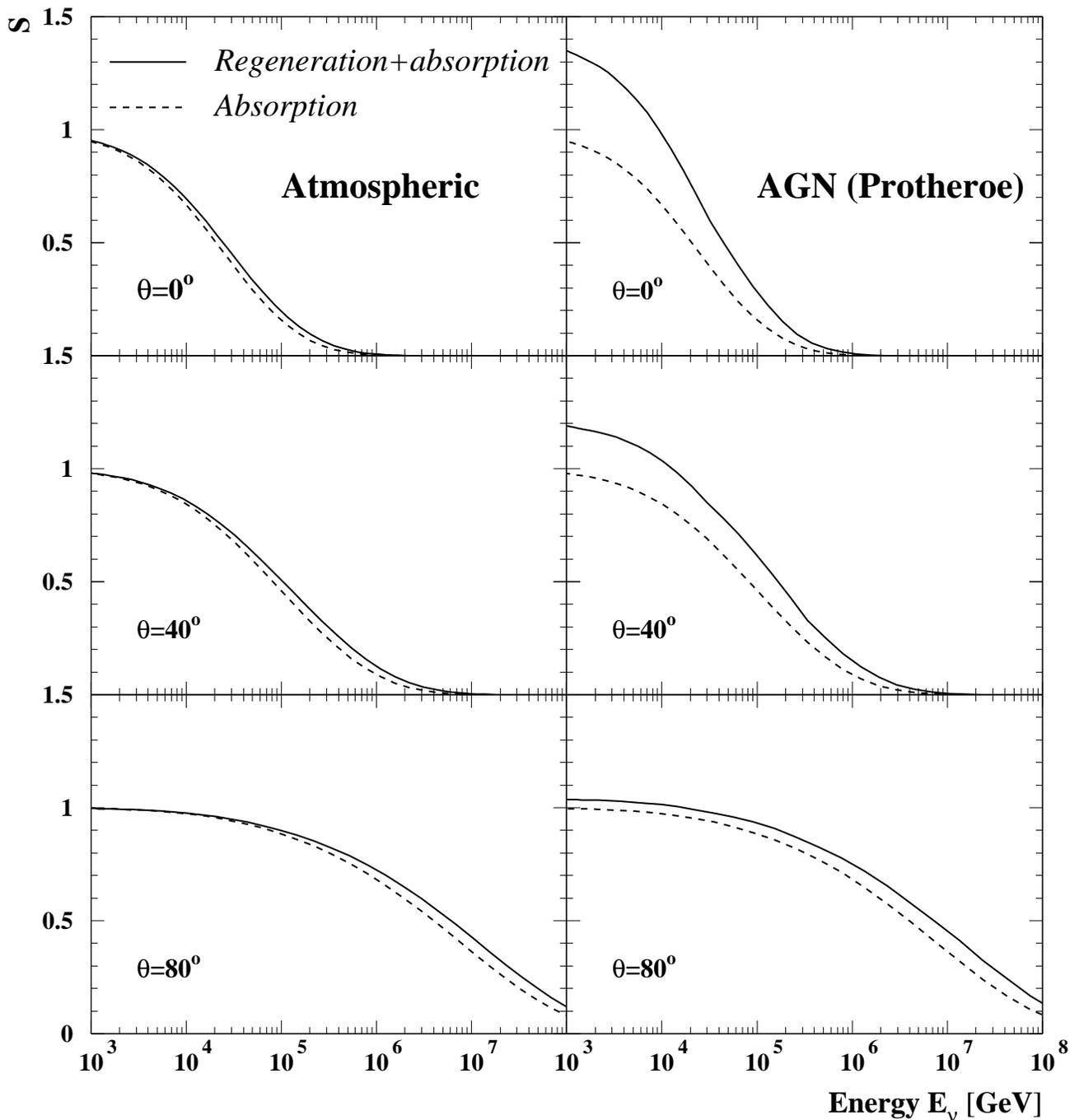
\item[Fig.~8] The shadowing factor $S$ of (\ref{eq:a21}) for two different initial                      
neutrino fluxes incident at three different nadir angles on a detector.  The angle $\theta                      
= 0^\circ$ corresponds to penetration right through the Earth's diameter.  The two                      
curves on each plot show the shadowing factor with and without NC regeneration                      
included.                     
                     
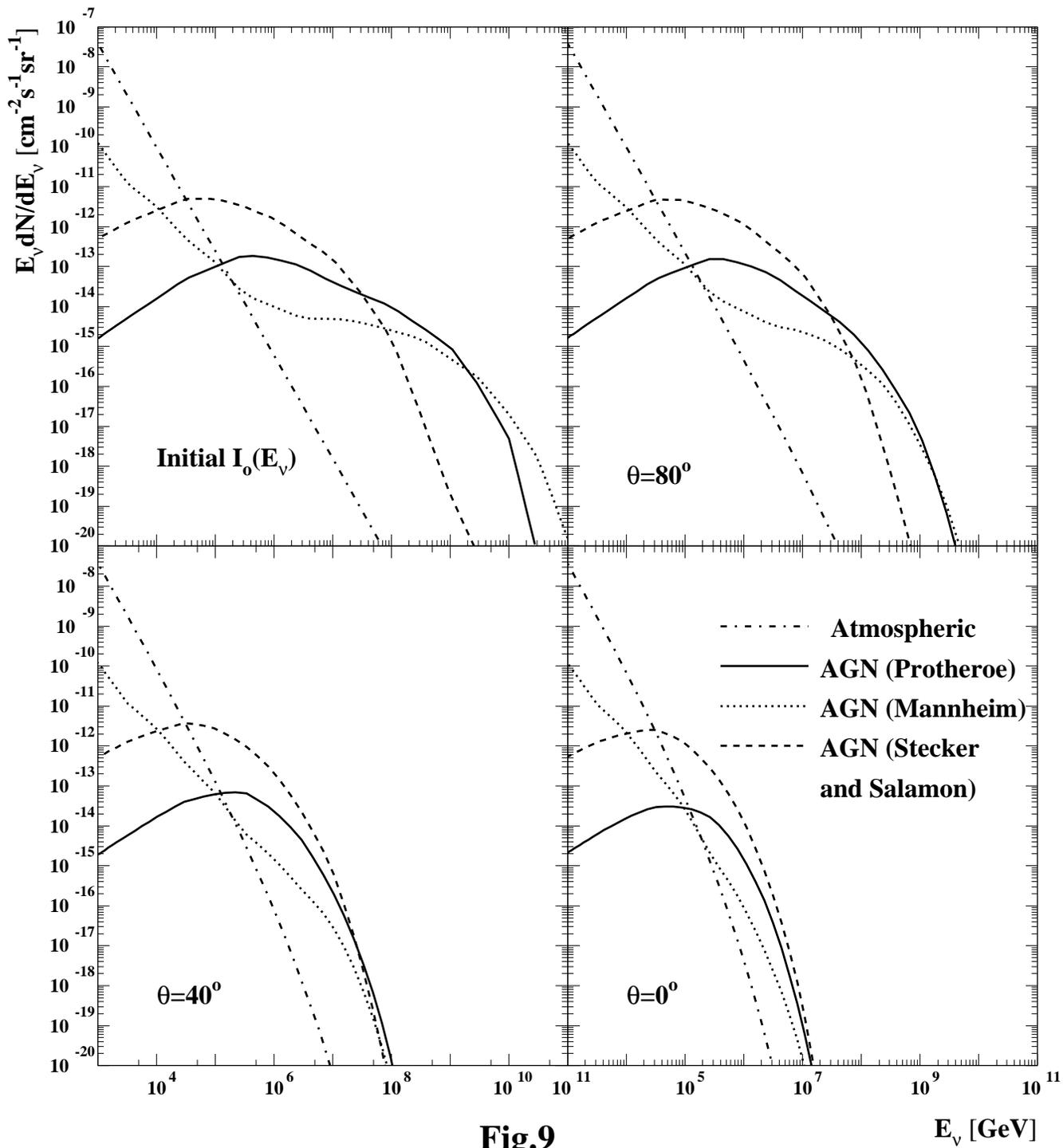
\item[Fig.~9] The initial flux $I_0 (E)$ and the flux at the detector $I (E)$ for three                      
different nadir angles corresponding to three models for AGN neutrinos                 
\cite{AGN1,AGN2,AGN3}.  The background atmospheric neutrino flux is also                 
shown.  All the fluxes are given for muon neutrinos. The corresponding fluxes 
from \cite{AGN1,AGN2,AGN3} were given originally for muon neutrinos
and anti-neutrinos, and their value has been divided by factor 2.                  
                     
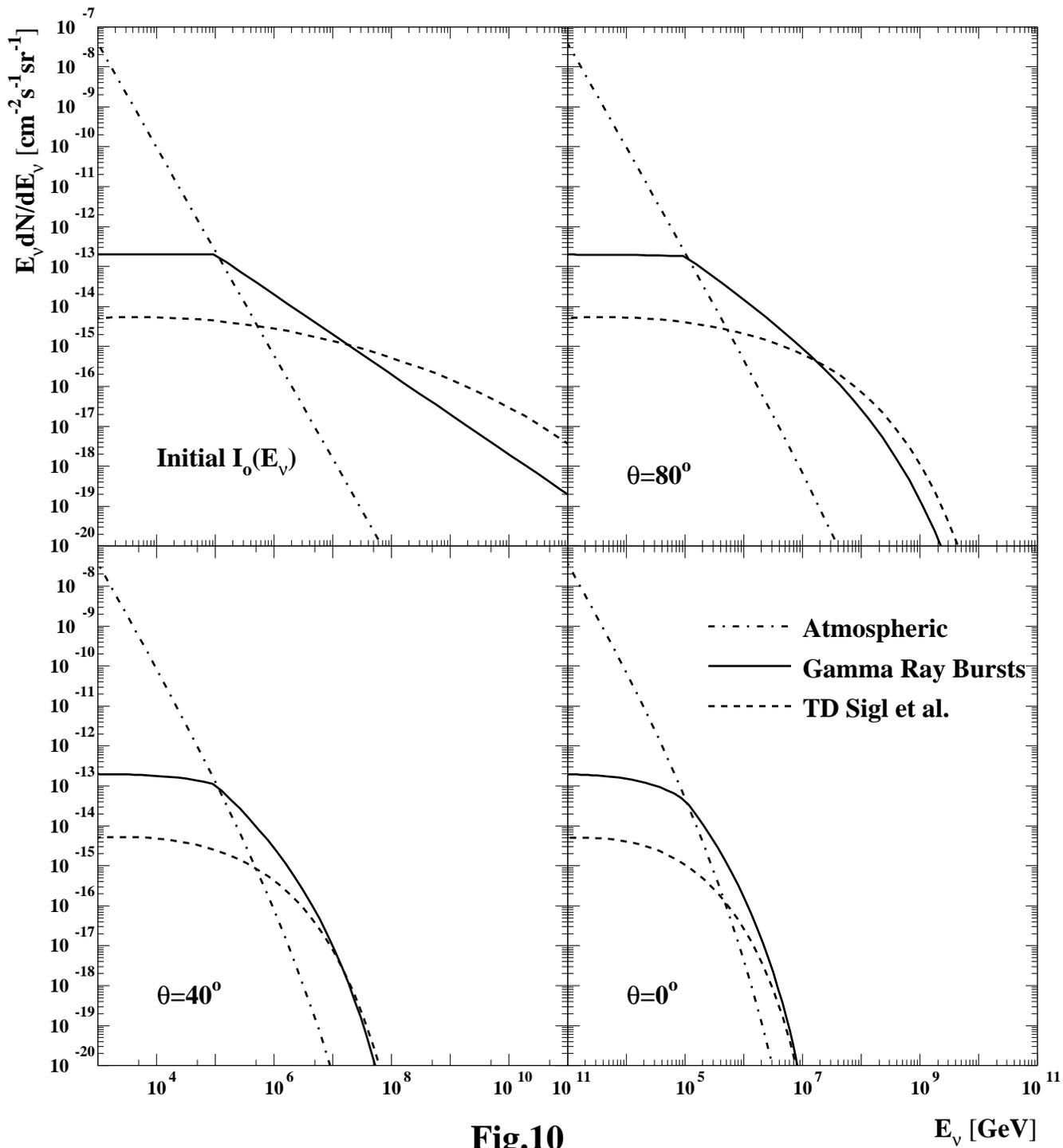
\item[Fig.~10] As for Fig.~9, but showing neutrino fluxes from gamma ray bursts                      
\cite{GRB} and from a top-down model \cite{TDM}.All the fluxes are given for muon neutrinos. The corresponding fluxes 
from \cite{GRB,TDM} were given originally for muon neutrinos
and anti-neutrinos, and their value has been divided by factor 2.                       
\end{itemize}                     
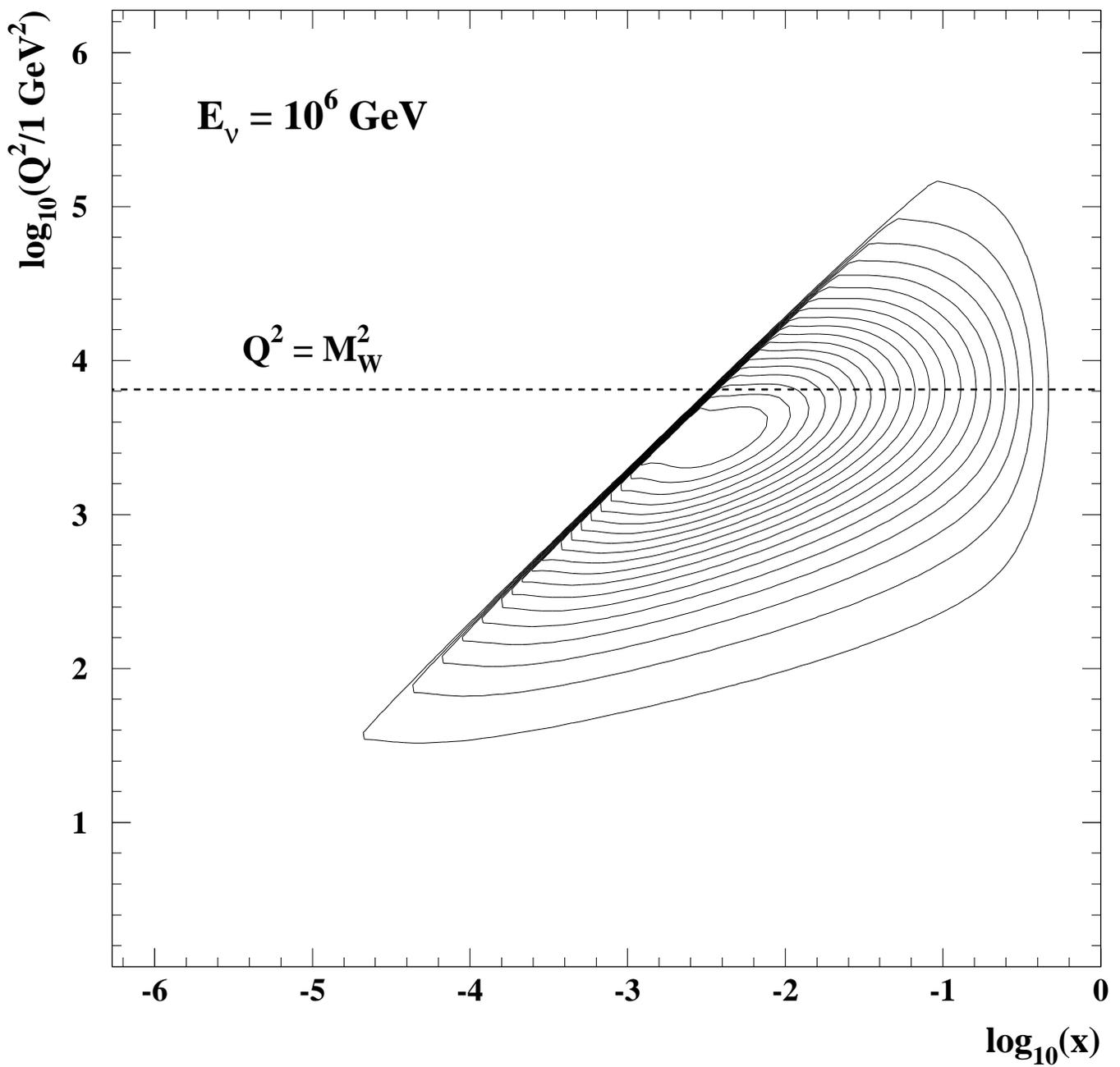
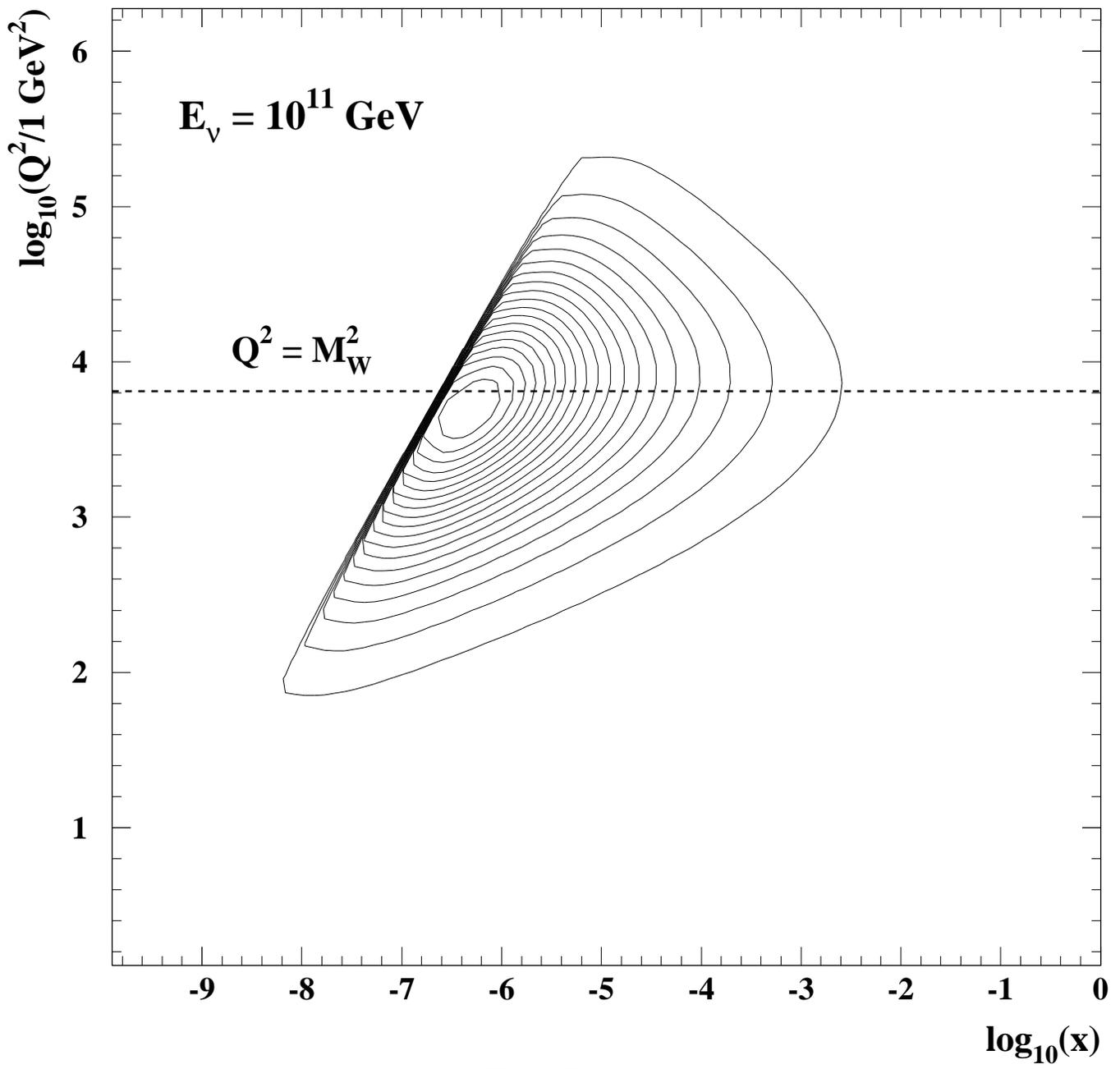
\end{document}